\documentclass[prl, 10pt,twocolumn, floatfix, groupedaddress, aps, longbibliography, nofootinbib]{revtex4-2}

\usepackage{times, mathrsfs, amsmath, amsfonts, graphics, graphicx, cancel, color, amsthm, bbm, mathtools, amssymb, physics}

\usepackage{amsmath,amsthm,amsfonts,graphicx,xcolor,times,xfrac,booktabs,mathtools,enumitem,xr,subfigure,amssymb,bbm,verbatim,appendix,placeins, physics}
\usepackage[unicode=true,bookmarks=true,bookmarksnumbered=false,bookmarksopen=false,breaklinks=false,pdfborder={0 0 1}, backref=false,colorlinks=true,citecolor=red]{hyperref}
\usepackage{orcidlink}

\usepackage{bm}
\usepackage{dcolumn,algorithm,algpseudocode}

\makeatletter
\theoremstyle{plain}

\theoremstyle{plain}

\theoremstyle{plain}

\theoremstyle{remark}
\newtheorem*{rem*}{\protect\remarkname}
\theoremstyle{plain}

\theoremstyle{plain}

\theoremstyle{definition}

\theoremstyle{plain}
\newtheorem*{thm*}{\protect\theoremname}
\theoremstyle{plain}
\newtheorem*{lem*}{\protect\lemmaname}

\providecommand{\propositionname}{Proposition}
\providecommand{\theoremname}{Theorem}
\providecommand{\lemmaname}{Lemma}
\providecommand{\remarkname}{Remark}
\providecommand{\conjecturename}{Conjecture}
\providecommand{\definitionname}{Definition}
\providecommand{\corollaryname}{Corollary}
\allowdisplaybreaks

\def\ket#1{\vert{#1}\rangle}

\def\BraVert{e.g.,roup\,\mid\,\bgroup}

\def\kket#1{\ket{#1}\!\rangle}
\def\bbkk#1#2{\langle\!\langle{#1}\vert{#2}\rangle\!\rangle}
\def\kkbb#1#2{\vert{#1}\rangle\!\rangle\langle\!\langle{#2}\vert}

\def\Tr#1{\mbox{Tr}\left[{#1}\right]}
\def\Var#1{\mbox{Var}\left[{#1}\right]}

\newcommand{\Id}{\mathbbm{1}}

\def\HiLi{\leavevmode\rlap{\hbox to \hsize{\color{yellow!50}\leaders\hrule height .8\baselineskip depth .5ex\hfill}}}

\begin{document}

\title{Optimizing quantum tomography via shadow inversion}
\author{Andrea Caprotti\orcidlink{0000-0002-4404-2216}}
\email{andrea.caprotti@univie.ac.at}
\affiliation{University of Vienna, Faculty of Physics, Vienna Center for Quantum Science and Technology (VCQ), Boltzmanngasse 5, 1090 Vienna, Austria}
\affiliation{University of Vienna, Vienna Doctoral School in Physics, Boltzmanngasse 5, 1090 Vienna, Austria}

\author{Joshua Morris\orcidlink{0000-0002-1022-7976}}
\email{joshua.morris@univie.ac.at}
\affiliation{University of Vienna, Faculty of Physics, Vienna Center for Quantum Science and Technology (VCQ), Boltzmanngasse 5, 1090 Vienna, Austria}
\affiliation{University of Vienna, Vienna Doctoral School in Physics, Boltzmanngasse 5, 1090 Vienna, Austria}

\author{Borivoje Daki\' c\orcidlink{0000-0001-9895-4889}}
\email{borivoje.dakic@univie.ac.at}
\affiliation{University of Vienna, Faculty of Physics, Vienna Center for Quantum Science and Technology (VCQ), Boltzmanngasse 5, 1090 Vienna, Austria}
\affiliation{Institute for Quantum Optics and Quantum Information (IQOQI), Austrian Academy of Sciences, Boltzmanngasse 3, 1090 Vienna, Austria.}

\begin{abstract}
In quantum information theory, the accurate estimation of observables is pivotal for quantum information processing, playing a crucial role in computational and communication protocols. This work introduces a  technique for estimating such objects, leveraging an underutilized resource in the inversion map of classical shadows that greatly refines the estimation cost of target observables without incurring any additional overhead. A generalized framework for computing and optimizing additional degrees of freedom in the homogeneous space of the shadow inversion is given that may be adapted to a variety of near-term problems. In the special case of local measurement strategies we show feasible optimization leading to an exponential separation in sample complexity versus the standard approach and, in an exceptional case, we give nontrivial examples of optimized postprocessing for local measurements, achieving the same efficiency as the global Cliffords shadows.
\end{abstract}

\maketitle
\section*{Introduction}

Quantum state tomography with classical shadows~\cite{ preskill2020}, initially motivated by the seminal work of Aaronson~\cite{aaronson17}, represents a significant advancement in the realm of quantum information processing, particularly in the efficient characterization of quantum systems.
 This technique, developed as a solution to the prohibitive resource requirements of full quantum state tomography, has since seen a plethora of modifications~\cite{hong-ye2023, vermersch2020,preskill2021,kraus2021,brierly2021,flammia2021,mezzacapo2022,miyake2021,jiang2023,babbush2023,bryan2024,bu2024,benedetti2023,bertoni2023,koh2022,fava2023,taylor2023,albert2024,nakaji2023,goldstein2022,guehne2023}, which all seek to provide a viable alternative for obtaining meaningful information about quantum states with substantially reduced computational and experimental overhead~\cite{serbyn2022, anshu2021, preskill2023, guehne2022, lu2021, kulik2021, borregaard2024, khemani2023, dakic2022, glos2022, ringbauer2022, vermersch2023}. 
Naturally, the main focus has been the ``quantum'' element of the procedure as this is where most practical efforts are stymied, either in gate or sample complexity. 
Other than machine learning \cite{preskill2023,mcclean2022} and adaptive techniques~\cite{maniscalco2021,glos2022}, little thought has been put towards explicit optimization of the classical postprocessing, with it often being treated as a mostly fixed step. 

At the core of shadow tomography lies the concept of constructing these aforementioned shadows, i.e., efficiently representable compressed classical descriptions of quantum states. 
These shadows are generated through a process involving random measurements on copies of a state, followed by classical processing to reconstruct a succinct representation of the state, the ``shadow''. 
These classical shadows, though not providing a complete description of the quantum state, contain enough information to estimate a wide range of properties with high accuracy when an appropriate Positive Operator-Valued Measurement (POVM) is used. 
An archetypal example is the set of measurements performed by random perturbations sampled from the Clifford group~\cite{webb2016,zhu2017} followed by a computational basis measurement
\begin{equation}
    \mathcal{C}_N = \{T \in SU(2^N)| T P_N T^\dagger \in \mathcal{P}_N \}/U(1),
\end{equation}
for all $P_N$, the set of Pauli operators on $N$ qubits. Critically for shadow tomography, we have that the Clifford group constitutes a $2,3$-design \cite{webb2016}, i.e., it can reproduce the statistical properties of the full set of unitaries up to the second and third moments. 
It is this that forms the basis upon which shadow tomography derives its impressive performance, implementing a map whose consequent inverse lies at the core of efficient estimation with classical shadows. 
If $U_k$ are elements of the $N-$qubit Clifford group and $\Pi_\ell$ are computational measurement projectors all acting on an input $N-$qubit quantum state then   
\begin{equation}\label{eq:true_twirl}
    \mathcal{M}[\rho] = \frac{1}{|\mathcal{C}_N|}\sum_{k=1}^{|\mathcal{C}_N|} \sum_{\ell=1}^{2^N}  T_{k\ell} \rho T_{k\ell} =  \lambda\rho  + (1-\lambda)\tr(\rho) \frac{\Id}{2^N},
\end{equation}
for $T_{k\ell} = U_k^\dagger \Pi_\ell U_k $. The inverse operation $\mathcal{M}^{-1}$ may be easily computed to recover the input state $\rho$ from the output $\mathcal{M}^{-1}$. Classical shadows' great insight was to show that this inversion procedure remains effective even when only a tiny fraction of Clifford measurements is used in the above sum.  This rapid convergence to a known configuration means the linear map of Eq.~\eqref{eq:true_twirl} may be readily inverted to recover the original state $\rho$. 
As shown in \cite{preskill2020}, the inverse map of Eq.~\eqref{eq:true_twirl} for global Clifford measurements on $N$ qubits seems to easily appear as the inverse depolarizing channel: $\mathcal{M}_N^{-1}(X) = (2^N+1)X-\Tr{X}\Id_{2^N}$; in the case of tensor products of random single-qubit Pauli measurements, the corresponding inverted quantum channel reduces to $\mathcal{M}^{-1}_\mathcal{P} = \bigotimes_{i=1}^N \mathcal{M}_1^{-1}$. 

\begin{figure*}[ht]
\includegraphics[width=0.98\textwidth]{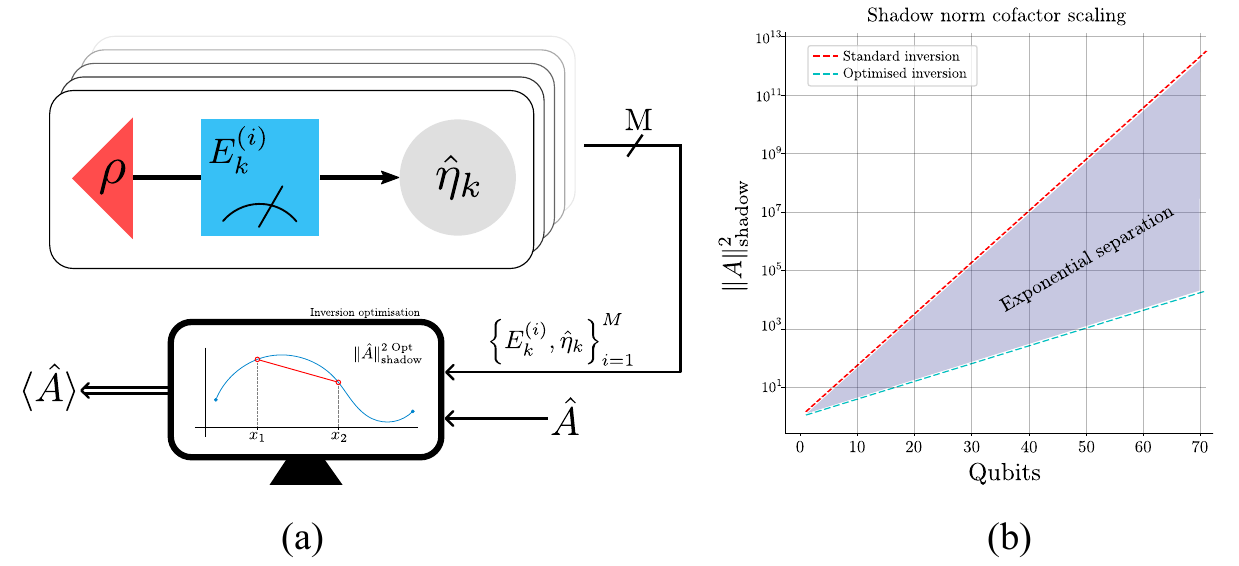}
\caption{(a) \textbf{Classical shadow tomography with  optimized shadow inversion.} 
After a collection of $M$ POVM outcomes has been acquired from a fixed quantum state, any estimation can be improved for a dynamic choice of target observable $A$ by computing an optimal ``inverse'' operation for that observable and POVM. 
(b) \textbf{Shadow norm for different strategies.} The shadow norm cofactor when optimized for certain target projectors, using local Pauli measurements on single qubits. 
The shadow norm for such an observable on $N$ qubits is normally $(3/2)^N$, but when optimized over additional degrees of freedom, it can be shrunk to $(1.15)^N$. 
Though still exponential in the number of qubits, the optimized strategy is exponentially cheaper than standard classical shadow tomography with local measurements.}
\label{fig:invmap}
\end{figure*}

In these cases, the inversion map takes a particularly simple form, and the estimation of consumed resources (sampling complexity) critically depends on it. 
Nevertheless, what is often neglected in studies is the fact that such a map is not unique, i.e., as long as the POVM that is used is overcomplete (in the operator, Hilbert-Schmidt space), an additional set of free parameters appears in the inversion process which may be optimized to improve the postprocessing performance of the estimation procedure. 
Surprisingly, this has been only recently recognized in the literature \cite{innocenti2023} by employing the techniques of measurement dual frames. 
In this work, we demonstrate that such additional degrees of freedom reside in the homogeneous space, i.e., the general inverse map contains a free homogeneous component, $\mathcal{M}^{-1}=\mathcal{M}_p^{-1}+\mathcal{M}_h^{-1}$, which can be further optimized. 
Our goal is to pursue such an optimization by considering a heretofore missed resource in the tomography procedure that appears most vividly in classical shadows and the postprocessing thereof. Though the natural choice of the inverse map (such as the one given above) is the inverse of the forward channel being applied, there is nothing that says it must be so. It is this inversion process, normally treated as a fixed mathematical procedure, that we aim to modify and improve upon. As long as the set of performed measurements is tomographically overcomplete, the additional degrees of freedom may be exploited to reduce case-specific variance for a given observable. In doing so, we will show that an exponential improvement in estimations is possible compared to the standard choice of inverse map presented in the literature. 
With this approach to processing acquired quantum data, we open a route to exploit areas where advantages may be gained without incurring an unacceptable resource cost.

\section*{Shadow map inversion}

We begin by describing a generalized procedure~\cite{morris2020, guehne2022}, depicted in Fig.~\ref{fig:invmap}(a), fulfilling the random measurement scheme with a fixed POVM $\lbrace E_k \rbrace_{k=1}^{n}$ on a $d$-dimensional system. 
The POVM elements randomly sample an unknown state $\rho$ with probability $p_k = \Tr{\rho E_k}$. 
These POVM elements span a subspace $\mathcal{V}_D=\mathrm{Span}\lbrace E_k \rbrace_{k=1}^{n}$ in the Hilbert-Schmidt space (HS) of the dimension $|\mathcal{V}_D|=D$. 
Typically, $D=d^2$ (the entire HS space), but for the sake of generality, we shall keep $D\leq d^2$. Furthermore, we assume $D<n$, and thus $E_k$ forms an overcomplete basis in $\mathcal{V}_D$. 
The estimation task aims to approximate the mean values of an observable $A\in\mathcal{V}_D$, i.e., $\ev{A}=\Tr{A\rho}$. 
Note that $\ev{A}=\Tr{A\bar{\rho}}$, with $\bar{\rho}$ being the projection of a quantum state $\rho$ onto $\mathcal{V}_D$ which follows directly from $A\in\mathcal{V}_D$. 
For the case of the entire HS space, we have $\bar{\rho}=\rho$. 
Similarly, the measured probabilities trivially satisfy $p_k=\Tr{E_k\rho}=\Tr{E_k\bar{\rho}}$.
The postprocessing task assigns a (matrix) estimator $\hat{\eta}\in\{\dots,\hat{\eta}_k,\dots\}$, i.e., for each outcome $E_k$, the corresponding matrix $\hat{\eta}_k$, which represents a single-shot approximation of $\bar{\rho}$. 
They can be calculated via an inversion map $\hat{\eta}_k=\mathcal{M}^{-1}(E_k)$~\cite{preskill2020,guehne2022}, which we will show is not unique, i.e., $\mathcal{M}^{-1}=\mathcal{M}_p^{-1}+\mathcal{M}_h^{-1}$, with the latter (homogeneous) component containing auxiliary parameters that may be used as a free resource in postprocessing.

The overall classical shadow itself is obtained by averaging over the estimators obtained from $K$ random measurements: $K^{-1} \sum_{s=1}^{K} \hat{\eta}^{(s)}$. 
Assuming that all samples are independent and identically distributed, the average value $\ev{\,\hat{\eta}\,}$ is immediately
\begin{equation}\label{eq:avgshadow_state}
    \ev{\,\hat{\eta}\,} = \sum_{k=1}^n p_k \hat{\eta}_k 
                      = \sum_{k=1}^n \Tr{\bar{\rho} E_k}\hat{\eta}_k 
                      = \bar{\rho},~~~\forall\bar{\rho}\in\mathcal{V}_D.
\end{equation}
This is the requirement on the estimator $\hat{\eta}$ such that the statistics of the shadow mimics, in expectation, that of the actual unknown state. 
With Eq.~\eqref{eq:avgshadow_state}, the task of finding an inversion map can be made equivalent to identifying a dual frame to the POVM in space of operators~\cite{innocenti2023}.
Since the POVM forms an overcomplete basis in $\mathcal{V}_D$, the solution to Eq.~\eqref{eq:avgshadow_state} is not unique. 
This multiplicity results in a set of free parameters, which becomes an extra resource that may be used in the postprocessing component of an estimation to reduce the variance.

To achieve this, we now show a constructive method to recover families of equivalent classical shadows in terms of the POVM elements and their free parameters. 
We rely on an equivalent problem to Eq.~\eqref{eq:avgshadow_state}, that is, to impose the equivalence between the expectation value of an observable $A$ and the corresponding classical estimator $a=\Tr{A  \hat{\eta}}$,
\begin{equation} 
    \ev{\,a\,} 
    = \Tr{A \ev{\hat{\eta}}} = \sum_{k=1}^n  p_k \Tr{A \,\hat{\eta}_k }
    = \Tr{\bar{\rho} A }. \label{eq:expval}
\end{equation}

It is instructive to express the operators as vectors in $\mathcal{V}_D$, i.e., for an operator $X\in\mathcal{V}_D$, we shall use the ket notation $\kket{X}$. 
We also identify an orthonormal basis $\beta_D=\lbrace\kket{B_j}\rbrace_{j=1}^D$ in $\mathcal{V}_D$.

The expectation value from Eq.~\eqref{eq:expval} can be expressed as
\begin{equation}\label{eq:expval_reconstruction}
    \ev{\,a\,} 
    = \sum_{k=1}^n  \bbkk{A ^\dagger}{\eta_k} \bbkk{E_k^\dagger}{\rho}
    = \bbkk{A ^\dagger}{\bar{\rho}}
\end{equation}
where $\bbkk{X}{Y}=\Tr{X^\dagger Y}$ is the inner product in $\mathcal{V}_D$.

To guarantee the estimators are unbiased, the requirement reduces to a resolution of the identity in the subspace $\mathcal{V}_D$,
\begin{equation}\label{eq:id_res_req}
    \sum_{k=1}^n \kkbb{\eta_k}{E_k^\dagger}=\sum_{k=1}^n \kkbb{\eta_k}{E_k} \equiv \Id_{D}, 
\end{equation}
which can be solved for $\kket{\eta_k}$ using $E_k^{\dagger}=E_k$. The last equation expressed in the basis $\beta_D$ reads
\begin{equation}\label{eq:req_coeff_matrix}
    \sum_{k=1}^n \bbkk{B_s}{\hat{\eta}_k}\bbkk{E_k}{B_t}
    = \sum_{k=1}^n L_{sk}R_{tk}^* = \delta_{st},
\end{equation}
with $L_{sk}=\bbkk{B_s}{\hat{\eta}_k}$ and $R_{tk}=\bbkk{B_t}{E_k}$. 
This implies the matrix equation $L R^{\dagger}=\Id_{D}$; thus, $L$ is a generalized inverse of $R^{\dagger}$. Since $R$ is a $D\times n$ matrix having $\kket{E_k}$ as columns, it is a rank-$D$ matrix (recall there is only a number $D$ of independent $E_k$ matrices). 
We can write its singular value decomposition (SVD), i.e., $R=U(\Sigma|\hat{0}) W^{\dagger}$ with $\Sigma=\mathrm{diag}[\sigma_1,\dots,\sigma_D]$ ($\sigma_s>0$) and $\hat{0}$ being the $D\times (n-D)$ zero matrix. 
Having this, a simple exercise shows $L=U(\Sigma^{-1}|H) W^\dagger$, with $H$ being an arbitrary $D\times (n-D)$ matrix.
Accordingly, we can split the solution into two parts, $L=L^{(p)}+L^{(h)}$, with $L^{(p)}=U(\Sigma^{-1}|\hat{0}) W^{\dagger}$ (particular) and $L^{(h)}=U(\hat{0}|H) W^{\dagger}$ (homogeneous).  
This results in a family of equivalent estimators,
\begin{equation}\label{eq:estimator_parthom}
    \hat{\eta}_k=\sum_{s=1}^D L_{sk}\hat{B}_s 
    =\sum_{s=1}^D \left( L_{sk}^{(p)}+L_{sk}^{(h)} \right) \hat{B}_s
    =\hat{\eta}_k^{(p)}+\hat{\eta}_k^{(h)}, 
\end{equation}
parameterized by a homogeneous solution $\hat{\eta}_k^{(h)}$. 
As we shall see, this multiplicity of possible inversion schemes represents an additional resource for error optimization in postprocessing.
We turn to observable properties which may be expressed in terms of estimator coefficients $a_k = \Tr{A  \hat{\eta}_k}=a_k^{(p)}+a_k^{(h)}$, with $a_k^{(p/h)}=\Tr{A  \hat{\eta}_k^{(p/h)}}$.
The homogeneous coefficients, from the definition of $\hat{\eta}_k^{(h)}$ in Eq.~\eqref{eq:estimator_parthom}, do not influence the expectation values or any other linear combination of classical shadows
\begin{equation}\label{eq:obs_reconstruction}
        \ev{\,a\,} =
        \sum_{k=1}^n p_k a_k =  
        \sum_{k=1}^n p_k \left(a_k^{(p)}+a_k^{(h)}\right) =  
        \sum_{k=1}^n p_k a_k^{(p)}.
\end{equation}
By substituting $p_k=\Tr{\rho\, E_k}$ and $\ev{\,a\,}=\Tr{A \,\rho}$ into the previous equation we find
\begin{align}
    & \sum_k\,a_k^{(p)}\,E_k=A,\label{eq:particular}\\
    & \sum_k\,a_k^{(h)}\,E_k=0. \label{eq:homogeneous}
\end{align}
The set of solutions to the latter (homogeneous) equation defines an auxiliary optimization space of dimension $n-D$ independent of the target observable (fixed by the choice of POVM). This is a free resource that may now be used in postprocessing.   
\begin{figure*}[ht]
\includegraphics[width=0.9\textwidth]{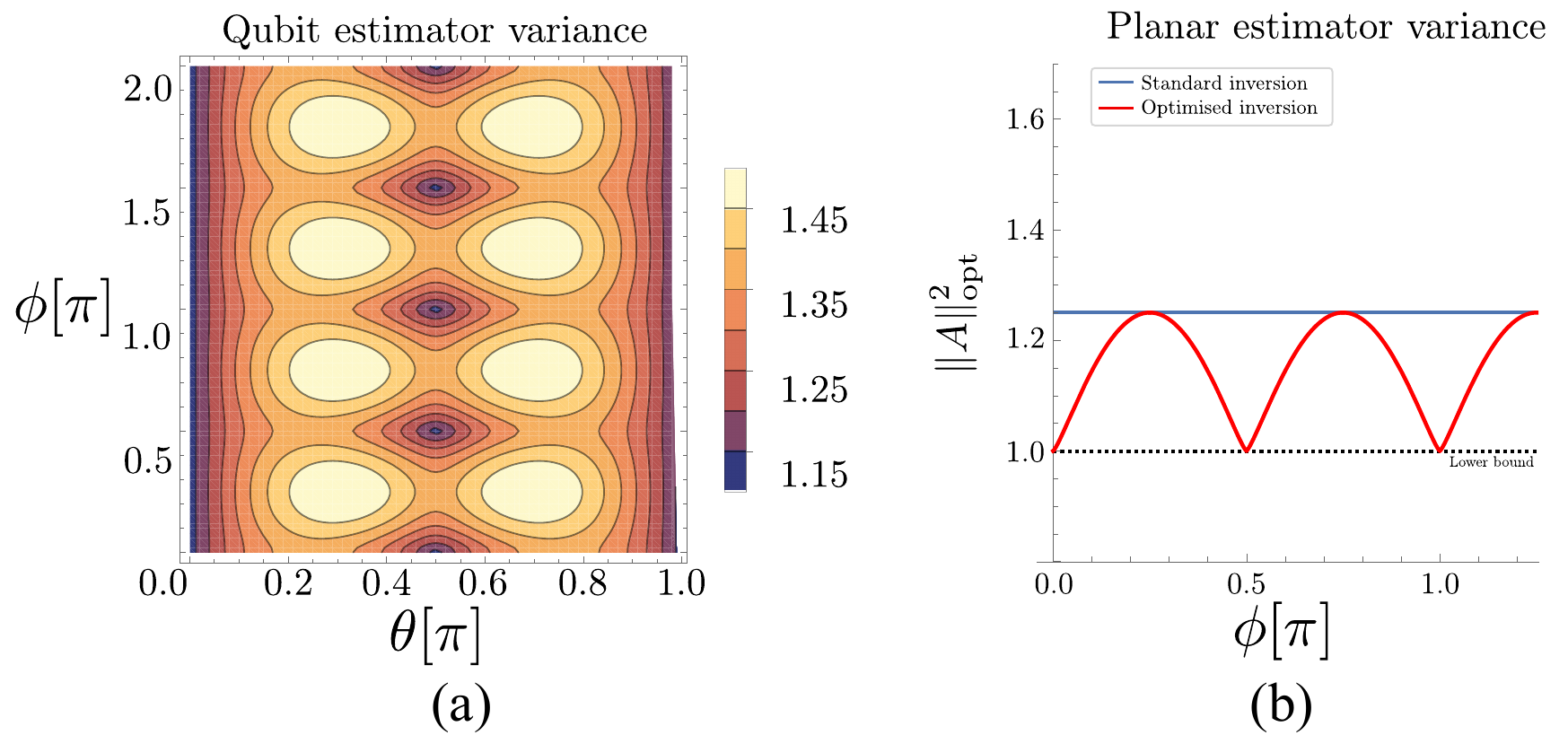}
\caption{ 
(a)  \textbf{Optimal free parameter for projectors on the Bloch sphere}. The contour plot shows the minimal shadow norm, given by Eq.~\eqref{eq:opt_shadow_norm}, of the single-qubit projector characterized by the Bloch sphere angles $\theta,\,\phi$ (in units of $\pi$).
Optimization results in bounds lower or equal than $3/2$, obtained via the Clifford inversion map, given by Eq.~\eqref{eq:pauli_estimators}, found in \cite{preskill2020}.
(b)  \textbf{Optimal free parameter for plane projectors}. 
The optimal value of the shadow norm of the single-qubit projector in the equatorial plane in terms of a single angle parameter $\phi$ (in units of $\pi$). 
This enhanced norm takes values in the range $1\leq\|\dyad{\psi}(\phi)\|^2_{\text{sh}}\leq 5/4$  (see main text).
In the case of $\phi = k \frac{\pi}{2},\,k\in\mathbb{Z}$, which corresponds to the projectors on the Pauli $X$ and $Y$ basis, the bound achieves optimal value  $\|\dyad{\psi}(\phi)\|^2_{\text{sh}}\equiv 1$.}
\label{fig:results}
\end{figure*}

Now, to effectively compare the different choices for estimators, we refer to the efficiency of the estimation process, that is, the expected sample size of copies required to estimate a property of the state up to a fixed accuracy. 
It is customary in classical shadow tomography to refer to variance as the figure of merit \cite{preskill2020}.
In particular, 
we consider the state-independent upper bound on variance, called \textit{shadow norm} $\|A\|^2_{\text{sh}}$, defined as
\begin{equation}\label{eq:shadow_norm}
\begin{split}
    \Var{a} &
    = \mathbb{E}[a^2]-(\mathbb{E}[a])^2 
    \leq \mathbb{E}[a^2] 
    = \sum_{k=1}^n p_k |a_k|^2
    \\
    &\leq 
    \max_{\sigma : \text{state}} 
            \Tr{\sigma \sum_{k=1}^n  |a_k^{(p)}+a_k^{(h)}|^2E_k}  = \|A\|^2_{\text{sh}} .
\end{split}
\end{equation}
Considering the upper bound to study sample complexity is reasonable since, for bounded observables, the quantity $(\mathbb{E}[a])^2$ is also bounded, and therefore the relevant contribution is brought by the first term $\mathbb{E}[a^2]$.

This quantity is equivalent to the maximal eigenvalue of the \textit{variance operator} defined as $\hat{O}_A \coloneq \sum_{k=1}^n |a_k^{(p)}+a_k^{(h)}|^2 E_k $; we will use both definitions interchangeably. 
This is akin to the original upper bound on the shadow norm and thence the sampling complexity found in \cite{preskill2020}, where the expected number of samples scales with $M=O(\varepsilon^{-2} \ln\,{\delta^{-1}} \|A\|^2_{\text{sh}})$. 
Nevertheless, this can be further optimized over the free parameters in the homogeneous part of the estimator.
By optimizing over these coefficients, it is possible to identify the configuration which guarantees the minimal sample size for the estimation of the particular observable $A$,
\begin{equation}\label{eq:opt_shadow_norm}
    \|A\|^{2}_{\text{opt}} =   
        \min_{a_k^{(h)} : H}
        \max_{\sigma : \text{state}} 
        \Tr{\sigma \sum_{k=1}^n
            |a_k^{(p)}+a_k^{(h)}|^2 E_k},
\end{equation}
The existence of a minimum is guaranteed through the positive semidefiniteness of the variance operator. We will provide examples to illustrate that different choices of the free parameters may result in a sampling complexity with exponential separation compared to that of the naive estimator construction, which shows, in turn, that optimization in Eq.~\eqref{eq:opt_shadow_norm} yields exceptional gains for relatively minuscule cost.

\section*{Product observables and shadow optimization}\label{sec:local_optimization}

Let us now discuss the estimation of product observables $A^{(N)}= A_1 \otimes A_2 \ldots \otimes A_N$ via local POVMs, i.e., of the type $E_{k_1}\otimes\dots\otimes E_{k_n}$. This case is particularly interesting for near-term applications, such as the Variational Quantum Eigensolver (VQE) algorithm \cite{peruzzo2014,tilly2022}. 
In such a scenario, it is easy to show that the shadow norm in Eq.~\eqref{eq:shadow_norm} has a product form (see,  also, \cite{guehne2022}) and, consequently, Eq.~\eqref{eq:opt_shadow_norm} becomes
\begin{equation}\label{eq:local_shadow_norm}
    \| A^{(N)} \|^{2}_{\text{opt}} 
    = \prod_{j=1}^N\| A_j \|^{2}_{\text{opt}}
\end{equation}
Since we are considering local measurements on independent subsystems, each term $\|A_j\|^2_{\text{opt}}$ can be optimized locally,  implying the optimization in Eq.~\eqref{eq:local_shadow_norm} is efficient overall. We see that even minimal deviation from the minimum in local shadow norms $\|A_j\|^2_{\text{opt}}$ may accumulate exponentially fast with $N$. This is illustrated in Fig.~\ref{fig:invmap}(b) for the case of the canonical inversion map of Eq. \eqref{eq:pauli_estimators}.
We now present examples of such exponential improvement for qubits for particular sets of local observables, specifically projectors, whereby by adjusting the available free parameters, we are able to reduce the expected bound on sample size based on assumptions for the observable itself.

\section*{Fidelity estimation of product states} \label{sec:proj_qubit}
We first consider the case of estimation of fidelity with product states. 
The target observables are thus tensor products of a projector on a single qubit $\bigotimes_{j=1}^N\dyad{\psi_j}$, parameterized on the Bloch sphere by two angles $(\theta_j,\,\phi_j)$.
The measurement scheme is described by the POVM of normalized projectors on the Pauli basis $E^{\pm}_{\zeta} = \frac{1}{6}(\Id \pm \sigma_{\zeta})$, where $\zeta = x,\,y,\,z$. Using the estimators 
\begin{equation}\label{eq:pauli_estimators}
    \hat{\eta}^{\pm}_{\zeta} = 3\left( 3E^{\pm}_{\zeta} -\Tr{E^{\pm}_{\zeta}}\Id_2 \right).
\end{equation}
found in \cite{preskill2020}, the shadow norm in Eq.~\eqref{eq:shadow_norm} simply evaluates for the generic single-qubit observable $A = a\cdot\Id + x\cdot\sigma_x + y\cdot\sigma_y+z\cdot\sigma_z$ to
\begin{equation}\label{eq:pauli_eigv}
    \|A \|^2_{\text{sh}}=
    a^2 + 3 (x^2 + y^2 + z^2) + 2|a| \sqrt{x^2 + y^2 + z^2}.
\end{equation}
Considering the particular case of projectors on the Bloch sphere, with the parameterization 
\begin{equation}\label{eq:sphere_coord}
    \begin{split}
        a \to \frac{1}{2}, 
        \quad \quad 
        x \to \frac{1}{2} \sin\theta \cos\phi, \\
        y \to \frac{1}{2}\sin\theta\sin\phi,
        \quad \quad
        z \to \frac{1}{2}\cos\theta ,
    \end{split}    
\end{equation}
 and Eq. \eqref{eq:pauli_eigv} trivially reduces to $\|\dyad{\psi}\|^2_{\text{sh}}\equiv\frac{3}{2}$.
Now we show how to improve this reference bound by our techniques. 
The POVMs $E^{\pm}_{\zeta}$ span the entire HS space and we may easily determine a homogeneous combination of the effects as in Eq.~\eqref{eq:homogeneous} given $\sum_{\zeta} h_{\zeta}^+ E_\zeta^+ +h_{\zeta}^- E_\zeta^-=0$.
Exploiting the symmetry of the system, they reduce to three parameters since $h_\zeta^+ = h_\zeta^-$:
\begin{equation}\label{eq:coef_hom_comb}
     \lbrace h_{\zeta}\rbrace_{\zeta = x,y,z} 
    = \lbrace p_1,p_2,-p_1-p_2 \rbrace
\end{equation}
As expected, we find $n-D=2$ parameters $p_1,p_2$.

Optimizing over these parameters, we present the minimal upper bound as a function of angles $(\theta, \phi)$ in Fig~.\ref{fig:results} (a).
The optimized results are contained within the interval $1.15 \leq \|\dyad{\psi_j}\|^2_{\text{sh}}\leq 1.5$.
Since the overall shadow norm given by Eq.~\eqref{eq:local_shadow_norm} is of the product form, there are entire regions in the $(\theta, \phi)$ space for which deviations from the standard $3/2$ lead to the exponential improvement in $N$.

\section*{Estimation with constrained POVMs}
In the previous section we have studied examples of overcomplete POVMs in the entire HS space, i.e. with $D=d^2$. 
Here we will pursue a relevant example of a POVM spanning only a subspace $\mathcal{V}_D$ of  lower dimension $D<d^2$,
in which $E_k$ forms an overcomplete basis (within $\mathcal{V}_D$).
Note that the standard inversion procedure \cite{preskill2020,guehne2022} is defined on the full HS space and does not directly apply in this instance.

Consider, for instance, the POVM constituted by the projectors of the $X$ and $Y$ Paulis $E^{\pm}_{\zeta} = \frac{1}{4}(\Id \pm \sigma_{\zeta})$, where $\zeta = x,\,y$, which spans the ``equator" on the Bloch sphere.
With a straightforward application of our procedure given by Eq.~\eqref{eq:estimator_parthom}, we obtain the following particular solution
\begin{equation}\label{eq:plane_pauli_estimators}
    \hat{\eta}^{\pm}_{\zeta} =  4 E^{\pm}_{\zeta} -\Tr{E^{\pm}_{\zeta}}\Id_{2}.
\end{equation}
The homogeneous term is added by solving
$\sum_{\zeta} h_{\zeta}^+ E_\zeta^+ +h_{\zeta}^- E_\zeta^-=0$,
where $\{h_\zeta\}_{\zeta=x,y} = \{ p, -p\}$ and $p$ is a free parameter .
For the generic planar observable (within the span of $\{E_\zeta^\pm\}$),
\begin{equation}\label{eq:plane_obs}
    A = \mqty(a & x+iy \\ x-iy & a),
\end{equation}
we find the shadow norm,
\begin{align}\label{eq:sn_pauli_plane}
    \|A \|^2_{\text{sh}} (p) 
    & = (a-x)^2 + 2(x^2+y^2) 
    + B(p) + \sqrt{C(p)}, \\
    B(p) &= \frac{1}{8}(2(p_r^2 +p_i^2) a^2+ (p_r(8+p_r)+p_i^2)x^2 \notag\\
    &+(p_r(p_r-8)+p_i^2)y^2), \notag\\
    C(p) &= x^2 (a (8 + p_r (6 + p_r) + p_i^2) - 
    2 (4 + p_r) x)^2 \notag\\
    &+ (a (8 + ( p_r-6) p_r + p_i^2) + 2 ( p_r-4) x)^2 y^2,
    \notag
\end{align}
where $p_r = \Re{p}$ and $p_i=\Im{p}$.

We can now proceed to two examples of 
estimation of product observables via local measurements achieving the same scaling of resources as global strategies~\cite{preskill2020}.

\subsection*{Optimal estimation of planar Paulis}
When considering, as the target observable, the planar Pauli,
\begin{equation}\label{eq:plane_pauli_coord}
        a \to 0 ,
        \quad
        x \to \cos\varphi,
        \quad
        y\to \sin\varphi,
\end{equation}
the optimized shadow norm from Eq.~\eqref{eq:sn_pauli_plane} results in $\|A\|^2_{\text{opt}} = 2$.
An immediate result is that the expected number of resources for the product of planar Pauli eigenstates scales as $2^N$, which matches the scaling obtained from global Clifford measurements~\cite{preskill2020}, $\|A\|^2_{\text{sh}} = \Tr{A^2} = 2^N$.
Surprisingly, we are able to match the optimal scaling~\cite{preskill2021_2} while still relying only on local single-qubit measurements.

\subsection*{Complete versus overcomplete POVMs}
In the context of the previous example, it is interesting to compare the performance of complete versus overcomplete POVM. 
For this purpose, consider the example of a triangle POVM on the plane, defined as the normalized projectors $E_k = \frac{2}{3} \dyad{\psi_k}$ of states 
\begin{equation}\label{eq:plane_triangle}
    \ket{\psi_k} = \dfrac{\ket{0}+e^{i\frac{k}{3}2\pi}\ket{1}}{\sqrt{2}} 
    \quad k=0,1,2
\end{equation}
Simple inversion of this complete basis brings us to estimators
\begin{equation}\label{eq:triangle_estimators}
    \hat{\eta}_k = 3 E_k -\frac{3}{4}\Tr{E_k} \Id_2.
\end{equation}
Since the POVM is complete, there is no free parameter dependence.
From here, we can compute the shadow norm as defined in Eq.~\eqref{eq:shadow_norm}.
For a planar Pauli as the target observable, as in Eq.~\eqref{eq:plane_pauli_coord}, the shadow norm is $\|A\|^2_{\text{sh}}=3,\,\forall \varphi$, which leads to an expected overhead of $3^N$ on the scaling of resources.
From the previous example, the same observables can be optimally estimated with an overhead of $2^N$.
We have, therefore, found an example in which an overcomplete POVM performs better than a complete measurement scheme in the same subspace.

\subsection*{Locally efficient fidelity estimation implies globally efficient}\label{sec:proj_plane}

We now present a configuration in which the optimization of the inversion map not only improves the sample complexity, but is actually able to achieve efficient estimation, i.e., $\|A\|^2_{\text{opt}} = 1$.

To show this, we choose, as target observables, tensor products of projectors that act on a qubit,
$A = \bigotimes_{j=1}^N \dyad{\psi(\phi_j)}$, where each single-qubit state is
parameterized as $\ket{\psi(\phi_j)}=(\ket{0}+e^{i\phi_j}\ket{1})\sqrt2$. 

Using the change of coordinates,
\begin{equation}\label{eq:plane_circ_coord}
    a \to \frac{1}{2} ,
    \quad 
    x \to \frac{1}{2} \cos\phi ,
    \quad
    y \to \frac{1}{2} \sin\phi,
\end{equation}
we express the generic shadow norm from Eq.~\eqref{eq:sn_pauli_plane} in terms of the parameter $\phi$,
\begin{align}\label{eq:sn_plane}
    \|\dyad{\psi}(\phi)\|^2_{\text{sh}}  =  
        \frac{1}{2}& (1 + p_i^2 + B (\phi, p_r) +\\ 
        &\quad\sqrt{2}\sqrt{ C (\phi, p_r)}) \notag\\
    B (\phi, p_r)  = (4&p_r-2)\cos{\phi} + \cos {2 \phi} +2 \notag\\
    &\quad+ 2 p_r (p_r-1)\notag \\
    C (\phi, p_r) = (4&p_r-1) \cos{\phi} + 2p_r\cos{2\phi}    \notag\\
    &\quad+\cos {3 \phi}+2 + 2 p_r (p_r-1),    \notag
\end{align}
where $p_r=\Re{p},\,p_i=\Im{p}$.
In Fig.~\ref{fig:results}(b), we present the optimal value of the shadow norm for $\phi\in [0,\pi]$.
We note that for each value of $\phi$, the optimized norm is $\|\dyad{\psi}(\phi)\|^2_{\text{opt}}\leq 5/4$.
For angles $\phi = 0,\pi/2,\pi,3 \pi/2$, the optimal shadow norm is exactly $\|E_\zeta^{\pm}\|^2_{\text{opt}} = 1$, corresponding to the eigenprojectors of $X$ and $Y$.

This is particularly relevant considering that X and Y are non-commuting observables, as it opens the possibility of the simultaneous estimation of up to $4^N$ projections simultaneously for complementary observables.
Previously, this result has been found in \cite{preskill2020} only via global measurement strategies relying on $N-$qubit Clifford unitaries. We have shown instead that this is achieved using exclusively local measurements.

\section*{Discussion}
In the course of this paper, we have observed how postprocessing, an often overlooked aspect in estimation, represents a relevant resource to be examined in the accounting of a more efficient tomography. 
By introducing a generalized framework to obtain families of equivalent parameter-dependent estimators, we have found a clear relation between overcomplete POVMs and their ability to reduce the variance in the estimation of arbitrary observables.
When considering local measurements on large composite quantum systems, even a small gain on each estimator represents an overall improvement in sample complexity, yielding an exponential separation between the naive sample variance and the optimized one as seen in Fig.~\ref{fig:invmap}(b). 
In the case of example two, this optimized gain can achieve a variance that performs as well as global classical shadows for given observables, with modest additional post processing that remains efficient. We anticipate that by considering overcomplete POVMs containing joint measurements on multiple qubits, further advantage may be found at increased (but still tractable) optimization difficulty, either in sample complexity or the set of observables. Finally, interleaving optimized multiqubit measurements may yield sample improvements that approach regular classical shadows within some limit. Such questions are natural candidates for further investigation and the identification of such a hierarchy of optimization will be the topic of future work. 

\textit{Note added} - Recently, we became aware of two other works \cite{malmi2024,fischer2024} considering a similar problem.
The authors consider a similar optimization of the postprocessing stage in estimation tasks, albeit referring to different frameworks.

\section*{Acknowledgements}
We thank Richard K\"ung for his insight and excellent discussions on the subject matter.
This research was funded in whole or in part by the Austrian Science Fund (FWF)(Grant No.10.55776/F71]) (BeyondC). 
%For open access purposes, the author has applied a CC BY public copyright license to any author accepted manuscript version arising from this submission.

\bibliography{bibliography}
\end{document}